\documentclass[12pt]{article}
\usepackage{amssymb}
\newcommand{\sect}[1]{\setcounter{equation}{0}\section{#1}}

\newcommand{\be}{\begin{equation}}
\newcommand{\ee}{\end{equation}}
\newcommand{\bea}{\begin{eqnarray}}
\newcommand{\eea}{\end{eqnarray}}
\newcommand{\beano}{\begin{eqnarray*}}
\newcommand{\eeano}{\end{eqnarray*}}
\newcommand{\nonu}{\nonumber \\}

\newcommand{\eps}{\epsilon}

\newcommand{\vph}{\varphi}

\newcommand{{\cf}}{\mbox{$\cal{F}$}}
\newcommand{{\cg}}{\mbox{$\cal{G}$}}
\newcommand{\ch}{\mbox{$\cal{H}$}}
\newcommand{\cj}{\mbox{${\cal J}$}}

\newcommand{\ct}{\mbox{$\cal{T}$}}
\newcommand{\cu}{\mbox{${\cal U}$}}
\newcommand{\cw}{\mbox{$\cal{W}$}}

\newcommand{\prt}{\partial}

\newcommand{\half}{\frac{1}{2}}

\newtheorem{defi}{Definition}
\newtheorem{theor}{Theorem}

\newcommand{\ie}{{\it i.e.}\ }
\newcommand{\CC}{\mbox{${\mathbb C}$}}
\newcommand{\RR}{\mbox{${\mathbb R}$}}

\newcommand{\ZZ}{\mbox{${\mathbb Z}$}}

\newcommand{\II}{\mbox{${\mathbb I}$}}

\newcommand{\bari}{{\bar{\imath}}}
\newcommand{\barj}{{\bar{\jmath}}}

\newcommand{\NP}[1]{Nucl.\ Phys.\ {\bf #1}}

\newcommand{\CMP}[1]{Comm.\ Math.\ Phys.\ {\bf #1}}

\newcommand{\LMP}[1]{Lett.\ Math.\ Phys.\ {\bf #1}}
\newcommand{\IJMP}[1]{Int. J.\ Mod.\ Phys.\ {\bf #1}}

\begin{document}
    \baselineskip=16pt
\renewcommand{\thefootnote}{\fnsymbol{footnote}}
\newpage
\pagestyle{empty}
\setcounter{page}{0}

\newcommand{\LAP}{LAPTH}
\def\logo{{\bf {\huge LAPTH}}}

\centerline{\logo}

\vspace {.3cm}

\centerline{{\bf{\it\Large 
Laboratoire d'Annecy-le-Vieux de Physique Th\'eorique}}}

\centerline{\rule{12cm}{.42mm}}

\vspace{20mm}

\begin{center}

  {\LARGE  {\sffamily Yangians and \cw-algebras}}\\[1cm]

\vspace{10mm}
  
{\Large C. Briot\footnote{briot@lapp.in2p3.fr}  
and
\underline{E. Ragoucy}\footnote{ragoucy@lapth.cnrs.fr}}\\[.42cm]
{\large Laboratoire de Physique Th{\'e}orique \LAP\\ 
     CNRS and Universit{\'e} de Savoie\\[.242cm]
     BP 110, F-74941  Annecy-le-Vieux Cedex, France. }
\end{center}
\vfill\vfill

\begin{abstract}
We present a connection between \cw-algebras and Yangians, in the case
of $gl(N)$ algebras, as well as for twisted Yangians and/or super-Yangians.
This connection allows to construct an $R$-matrix for the \cw-algebras, 
and to classify their finite-dimensional irreducible representations. 
We illustrate it in the framework of nonlinear Schr\"odinger
equation in 1+1 dimension.
\end{abstract}

\vfill
\begin{center}
This note corresponds to a talk given by E. Ragoucy at \\
{\it NEEDS 2000} (Gokova, Turkey, July 2000); \\
{\it Group23} (XXIII$^{th}$ ICGTMP, Dubna, Russia, August 2000).
\end{center}
\vfill
\rightline{\LAP-819/00-Conf}
\rightline{October 2000}

\newpage
\pagestyle{plain}
\setcounter{footnote}{0}

\sect{Introduction}
\cw-algebras have been introduced in the $2d$-conformal models as a 
tool for the study of these theories. Then, these algebras and their 
finite-dimensional versions appeared to be relevant in several 
physical backgrounds. However, a full understanding of their algebraic 
structure (and of their geometrical interpretation) is 
lacking. The connection of some  of these \cw-algebras with Yangians 
appears to be a solution at least for the algebraic structure: it allows 
the construction of an $R$-matrix for \cw-algebras, the classification of their 
irreducible finite-dimensional representations and the determination of their center.

The paper is structured as follows: in section \ref{yang} (section \ref{Walg}) 
we remind some basic definitions for Yangians (for \cw-algebras). 
Then, the connection between these two objects is presented in section \ref{YWglN} for
the case of $gl(N)$. Section \ref{NLS} is devoted to a physical example where the 
connection explicitly appears, namely the Nonlinear Schr\"odinger equation in 
1+1 dimension. The two following sections present various generalizations: the case of $so(M)$ and 
$sp(2M)$ algebras is studied in section \ref{twist}, and the case of superalgebras in section
\ref{super}. We conclude in section \ref{concl}.

For the sake of concision, 
we have chosen to detail, as an illustrative case, the study of $Y(N)\equiv Y(gl(N))$ and 
$\cw_{p}(N)\equiv\cw[gl(Np),N.sl(p)]$, while being less precise on 
the generalizations, sending the interested reader back to the original papers.

\sect{Yangian $Y(\cg)$\label{yang}}
Yangians $Y(\cg)$, associated to each simple Lie algebra \cg, have
been introduced by Drinfeld as deformation of (half) a loop algebra
based on \cg \cite{Drinfeld}. They have generators: 
\be
Y(\cg)=\cu\left(\rule{0ex}{2.1ex}Q^a_n,\ a=1,\dots\mbox{dim}(\cg);\ 
n=0,1,\ldots,\infty\right)
\ee
$n$ is the loop index and $a$ labels the \cg-adjoint representation. 
In other words, 
we have an infinite set of adjoint representations (labeled by $n$), the
first one being $\cg$ itself. This is gathered in the relations:
\be
{[Q_0^a,Q_n^b]}={f^{ab}}_c\, Q^c_n \label{AdjRep}
\ee
The deformation appears in the remaining relations:
\be
{[Q_m^a,Q_n^b]}={f^{ab}}_c\, Q^c_{m+n}+P^{ab}_{nm}(Q)
\ee
where $P^{ab}_{nm}$ is a polynomial in the $Q$'s.

Yangians are Hopf algebras, their coproduct being given by
\bea
\Delta(Q_0^a) &=& Q_0^a\otimes\II+\II\otimes Q_0^a\\
\Delta(Q_1^a) &=& Q_1^a\otimes\II+\II\otimes Q_1^a+\half{f^a}_{bc}\,
Q_0^b\otimes Q_0^c
\eea
which also shows the deformation with respect to the loop 
algebra coproduct. 

There is a consistency relation, which takes the form of a Jacobi-like 
identity. Depending on the algebra, this Jacobi-like identity takes 
the form (\ref{yG}) when $\cg\neq sl(2)$, or the form (\ref{eq:6}) 
when $\cg=sl(2)$:
\be
{f^{bc}}_d {[Q_1^a,Q_1^d]} + {f^{ca}}_d {[Q_1^b,Q_1^d]} 
+ {f^{ab}}_d {[Q_1^c,Q_1^d]}\  =\  
{f^a}_{pd}{f^b}_{qx}{f^c}_{ry}{f^{xyd}}
\, s_3(Q_0^p,Q_0^q,Q_0^r) \label{yG}
\ee
\be
\begin{array}{l}
{f^{cd}}_e {[[Q_1^a,Q_1^b],Q_1^e]} + 
{f^{ab}}_e {[[Q_1^c,Q_1^d],Q_1^e]}\  =\  \\
 \hspace{2.4em} \left(
{f^a}_{pe}{f^b}_{qx}{f^{cd}}_y{f^y}_{rz}{f^{xz}}_g+
{f^c}_{pe}{f^d}_{qx}{f^{ab}}_y{f^y}_{rz}{f^{xz}}_g\right)
 \eta^{eg}\ s_3(Q_0^p,Q_0^q,Q_1^r) \label{eq:6} 
 \end{array}
\ee
where $s_{3}$ is the symmetrized product.

In the case of $\cg=gl(N)$, Yangians admit an $R$-matrix presentation
\cite{YRTT,YN}:
gathering the generators into an $N\times N$ matrix, and using a spectral 
parameter $u$, one defines
\be
T(u)=\sum_{i,j=1}^N\sum_{n=0}^\infty\ u^{-n}T_{n}^{ij}E_{ij}=\sum_{i,j=1}^N
T^{ij}(u)E_{ij}
\ \mbox{ with }\ T_0^{ij}=\delta^{ij}
\ee
Then, the defining relations of $Y(gl(N))=Y(N)$ become
\bea
&& R_{12}(u-v)T_1(u)T_2(v)=T_2(v)T_1(u)R_{12}(u-v)\\
&& \Delta T(u)=T(u)\otimes T(u)\ ;\ S(T(u))=T(u)^{-1}\ ;\ \eps(T(u))=1
\eea
where 
\bea
&& R_{12}(x)=\II_N\otimes\II_N-\frac{1}{x}P_{12} \ \mbox{ with }\ 
P_{12}=\sum_{i,j=1}^N\, E_{ij}\otimes E_{ji} \label{Rmat}\\
&& T_1(u)=T(u)\otimes\II_N=\sum_{i,j=1}^N T^{ij}(u)\, 
E_{ij}\otimes\II_N\\
&& T_2(u)=\II_N\otimes T(u)=\sum_{i,j=1}^N T^{ij}(u)\, 
\II_N\otimes E_{ij}
\eea
$R$ is a rational solution to the Yang-Baxter equation, and $P_{12}$ is the 
permutation operator of the two auxiliary spaces (spanned by the 
$N\times N$ matrices).
\subsection{Classical Yangians\label{class}}
In the following, we will be interested mainly by a classical version
of the Yangians, where the commutators are replaced by Poisson brackets.

{For} the first presentation, the relations are the same, except for the Poisson 
bracket which now replaces the commutator. For instance, the relation
(\ref{yG}) becomes:
\be
{f^{bc}}_d \{Q_1^a,Q_1^d\} + {f^{ca}}_d \{Q_1^b,Q_1^d\} 
+ {f^{ab}}_d \{Q_1^c,Q_1^d\}\  =\  
{f^a}_{pd}{f^b}_{qx}{f^c}_{ry}{f^{xyd}}
\, Q_0^p Q_0^q Q_0^r
\ee

{For} $Y(N)$, the Poisson bracket appears as a classical version of the 
commutator:
\bea
&& R(x)=\II-\hbar\, r(x)\ ;\ [.,.]=\hbar\{.,.\}\ ;\ T(u)=L(u)\\
&& \{L_1(u),L_2(v)\} = [r_{12}(u-v),L_1(u)L_2(v)]\ ;\ r(x)=\frac{1}{x}P_{12}
\eea

\sect{$\cw(\cg,\ch)$ algebras\label{Walg}}
\cw-algebras have been first introduced in the context of $2d$-conformal 
theories by Zamolodchikov \cite{Zam} as a tool for
classifying the irreducible unitary representations of these theories. 
Since, they appeared to be symmetries of Toda field theories \cite{Oraf,Boscho}. 
In this context, \cw-algebras are constructed as Hamiltonian reduction 
of affine (Kac-Moody) algebras. Later on, a simpler version of these 
algebras, called finite \cw-algebras, was introduced by De Boer and Tjin 
\cite{dBTj}. They are constructed as Hamiltonian reduction of finite
dimensional Lie algebras: the resulting algebra is a polynomial algebra
with a finite number of generators. 

More precisely, starting from a Poisson-Lie algebra \cg, one constrains some of 
the generators of \cg. The constraints are second class, and one considers
the Dirac brackets deduced from these constraints: the \cw-algebra
is defined as the set of unconstrained generators provided with the 
Dirac brackets. The system of constraints is given by a subalgebra
\ch\ of \cg, whence the denomination \cw(\cg,\ch), 
see \cite{Oraf,Boscho,classW} for more details.

Here, we will be concerned with a class of finite \cw-algebras:
$\cw[gl(Np),N.sl(p)]$ algebras. We will denote these algebras $\cw_p(N)$. 
The generators of $\cw_p(N)$
are in finite number: 
\be
\cw_p(N)=\cu\left(W^a_m,\ a=1,2,\dots,N^2;\  n=1,2,\dots,p\right)
\ee
 They obey to
\be
\{W^a_0,W^b_n\} = {f^{ab}}_c\, W^c_n \ \mbox{ and }\
\{W^a_m,W^b_n\} = {f^{ab}}_c\, W^c_{m+n} +P^{ab}_{nm}(W)
\ee
where $P^{ab}_{nm}(W)$ are polynomials in the $W$ generators.

Its similarity with the 
Yangian presentation is quite appealing and has motivated the 
studies in this direction.

\sect{$Y(N)$ and $\cw_p(N)$\label{YWglN}}
{From} the previous presentations, it is natural to seek for a relation
between $\cw_p(N)$ algebras and Yangians $Y(N)$. Indeed, such
a relation exists, and it has been proven in \cite{RS}:
\begin{theor}
There is an algebra homomorphism between $\cw_p(N)$ algebras and 
Yangians $Y(N)$. More precisely, there is a one-to-one connection between
the first $p\,N^2$ generators of $Y(N)$ and the generators of the $\cw_p(N)$
algebra:
\be
Q_n^a\ \rightarrow\ \beta_n^a\, W_n^a+R_n^a(W)
\mbox{ with }\beta_n^a\in\RR\setminus\{0\}
\ee
$R_n^a(W)$ are polynomials in the $W_m^b$ with $m<n$.
The remaining generators of $Y(N)$ are polynomials in the \cw-generators.
\end{theor}
It has been proven that the generators of the \cw-algebra obey to the Jacobi-like
relations that define the Yangian.

The $R$-matrix approach is an easier way to tackle this relation \cite{wrtt}:
\begin{theor}\label{theo-wrtt}
$\cw_p(N)$ is isomorphic to the truncated Yangian $Y_p(N)$,
defined by
\[
Y_p(N)=Y(N)/\cj_p\mbox{ with }\cj_p\mbox{ ideal generated by }
\ct_p=\{T_n^{ij},\ i,j=1,\dots,N;\ n>p\}
\]
\end{theor}
Thanks to this theorem, one gets an $R$-matrix formulation of the 
$\cw_p(N)$ algebras:
\beano
&&\{W_1(u),W_2(v)\}=[r_{12}(u-v),W_1(u)W_2(v)]\\
&& \mbox{ with }W(u)=\sum_{i,j=1}^N\sum_{n=0}^p W_n^{ij} u^{-n} E_{ij}
\mbox{ and }r(x)=\frac{1}{x}P_{12}
\eeano

\null

{\bf Remark:} The Hopf structure of $Y(N)$ does {\underline{not}} survive the coset, so that
the algebra isomorphism of theorem \ref{theo-wrtt} is in this sense
 a no-go theorem about the
existence of a natural Hopf structure for \cw-algebras.

\null

One can also determine the center of the $\cw_p(N)$ algebra:
\begin{theor}
The center of the $\cw_p(N)$ has dimension $Np$ and is
canonically associated to the center of the underlying $gl(Np)$ algebra.
\end{theor}
Moreover, since the irreducible finite-dimensional representations of the Yangian
have been classified, one can prove:
\begin{theor}
All the finite dimensional irreducible representations of $\cw_p(N)$
are highest weight. They are in one-to-one correspondence 
with the families 
$\{P_1(u),\dots,P_{N-1}(u),\rho(u)\}$, where $P_i(u)$ are polynomials of
the form 
\be
P_i(u)=\prod_{k=1}^{d_i}(u-\gamma_i^k) \mbox{ with } \sum_i d_i\leq p
\mbox { and } \gamma_i^k\in\CC
\ee
 $\rho(u)=1+\sum_{n=1}^{Np}c_nu^{-n}$ codes the values $c_n$ 
of the Casimir operators in the representation.
\end{theor}
All these representations are highest weight, the highest weight being
reconstructed from the polynomials $P_i$ through:
\be
\frac{\mu^{i}(u)}{\mu^{i+1}(u)}=\frac{P_i(u+1)}{P_i(u)},\ i=1,\dots,N
\ee
with the highest weight vector $\xi$ defined by:
\be
W^{ii}(u)\xi = \mu^i(u)\xi \ \ 1\leq i\leq N\ \mbox{ and }\
W^{ij}(u)\xi = 0,\ \ 1\leq i<j\leq N
\ee

Finally, let us remark that a detailed analysis of the decomposition 
of $\cw(\cg,\ch)$ algebras with respect to their Lie subalgebras
 (using the technics developed in \cite{classW})
shows that such a connection cannot exist with Yangians $Y(\cg)$
when $\cg=so(N)$
or $sp(2N)$. Indeed, when $\cg$ is not a $gl(N)$ algebra, there is no
$\cw(\cg,\ch)$ algebra such that all its generators are in adjoint 
representations of the Lie subalgebra of $\cw(\cg,\ch)$.
We will see below that the connection apply to objects different
from the $Y(\cg)$ Yangians.

\sect{Nonlinear Schr\"odinger equation in 2 dimensions\label{NLS}}
The Nonlinear Schr\"odinger equation in two dimensions (NLS) is a nice framework
where the connection between $Y(N)$ and $\cw_p(N)$ can be visualized.

We start with
\be
i\prt_t\Phi = \prt_x^2\Phi+g\,|\Phi|^2\,\Phi\ \mbox{ with }\ 
\Phi^t=\left(\vph_1,\ldots,\vph_N\right)
\mbox{ and }g<0
\ee
The (quantum) solutions to this equation are known for a long time \cite{Ros}.
They take the form
\beano
\Phi &=& \sum_{n=0}^\infty g^n\, \Phi_{(n)}\ \ ;\ \ 
\Omega_n = q_0x-q_{0}^2t+
\sum_{i=1}^n\left(\rule{0ex}{2.4ex}(q_i-p_i)x-(q^2_i-p^2_i)t\right)
\\ 
\Phi_{(n)} &=& \int d^{n+1}q\, d^np\ 
a^\dag_1(p_1)\dots a^\dag_n(p_n)a_n(q_n)\dots a_0(q_0)\, 
\frac{\exp(i\Omega_n)}{\prod_{i=1}^n(p_i-q_{i-1})(p_i-q_i)}
\eeano
where the $a$'s and $a^\dag$'s obey a Zamolodchikov-Faddeev (ZZF) algebra
\cite{ZF}:
\bea
a_1(k_1)\, a_2(k_2) &=& R_{12}(k_2-k_1)\, a_2(k_2)\, a_1(k_1) \\
a^\dag_1(k_1)\, a^\dag_2(k_2) &=& a^\dag_2(k_2)\, a^\dag_1(k_1) \,
R_{12}(k_2-k_1)  \\
a_1(k_1)\, a^\dag_2(k_2) &=& a^\dag_2(k_2)\, R_{12}(k_1-k_2) \, a_1(k_1) 
+\delta_{12}(k_1-k_2)
\eea
where $R$ is the matrix of the Yangian $Y(N)$. We use the notation
\[
\begin{array}{ll}
 R_{12}(x)=R_{ik}^{jl}(x)\, E_{ij}\otimes E_{kl} &\\
a_1(k)=a_i(k)\, v_i\otimes\II & a_2(k)=a_i(k)\, \II\otimes v_i \\
a^\dag_1(k)=a_i(k)\, v^\dag_i\otimes\II &
a^\dag_2(k)=a_i(k)\, \II\otimes v^\dag_i \\
E_{ij}v_k=\delta_{jk}v_i\ ;\ v^\dag_k E_{ij}=\delta_{ik}v^\dag_j\ \ &
 v_i\cdot v^\dag_j=\delta_{ij}\ ;\ v^\dag_i\cdot v_j=E_{ij}
 \end{array}
\]
The apparition of the Yangian's $R$-matrix
is  not surprising in this context, since the Yangian is a symmetry
of NLS. Indeed, in \cite{NLS}, the generators of this algebra have been
expressed in term of the ZZF algebra. They take the form
\bea
Q_s^a &=& \sum_{n=0}^\infty \frac{(-1)^n}{n!} Q_{s,(n)}^a\ 
\mbox{ with }s=0,1
\label{Qas}\\
Q_{s,(n)}^a &=& \int d^nk\, a_1^\dag(k_1)\dots a_n^\dag(k_n)\ J^a_{s,(n)}\ 
a_n(k_n)\dots a_1(k_1)
\eea
where $J^a_{s,(n)}$ belongs to $M(N,\CC)^{\otimes n}(k_1,\dots,k_n)$, 
 $M(N,\CC)$ being the space of $N\times N$ matrices
 (see \cite{NLS} for the exact expression).
 The Yangian is a symmetry of the whole hierarchy associated to
NLS, as it can be seen from the expression of the Hamiltonians $H_m$ in 
terms of the ZZF algebra:
\be
H_m=\int dk\, k^m\, a^\dag(k)a(k)\ \ \Rightarrow\ \ [H_{m},Q_s^a]=0
\ee
In fact, the generators $a^\dag(k)$ correspond to the asymptotic states
of the NLS hierarchy, and it is natural to look at the Fock space $\cf$
spanned by the $a^\dag$'s. This Fock space naturally decomposes into
eigenspaces of the particle number $H_0$: $\cf=\oplus_p \cf_p$.

Now, on each subspace $\cf_p$, the sums (\ref{Qas}) truncate at level $n=p$,
in the same way one defines $\cw_p(N)$ from $Y(N)$. 

{\it Thus, on each
subspace $\cf_p$, the action of $Y(N)$ reduces to the $\cw_p(N)$ algebra.}

\sect{Orthogonal and symplectic cases\label{twist}}
\subsection{Folding $\cw_p(N)$}
It is known for a long time that $so(M)$ and $sp(2M)$ algebras can
be obtained from $gl(N)$ ones using their outer automorphism. If $\tau$
is such an automorphism, the $so(M)$ and $sp(2M)$ algebras are obtained as
Ker$(\II-\tau)$, \ie the algebra of $\tau$-invariant generators of $gl(N)$.

It is the same technics that is used for \cw-algebras. Indeed,
it has already been showed that \cw-algebras based on $so(M)$ and $sp(2M)$
can be constructed from the ones based on $gl(N)$ \cite{fold}. 
The Hamiltonian reduction
(\ie the constraints) must be compatible with the folding (\ie the automorphism)
so that not all the $\cw(gl(N),\ch)$ algebras can be folded. However, it is
enough to produce all the \cw-algebras based on $so(M)$ and $sp(2M)$. 

Here, we will consider only the folding of $\cw_p(N)$, which 
can indeed be folded. 
The automorphism we consider has been defined in \cite{Ytw}. It takes
the form:
\be
\hspace{-1.2ex}
\tau_\pm(W^{ij}_n)=(-1)^{n+1}\, \theta^i\theta^j\, W^{N+1-j,N+1-i}_n
\ \left\{\begin{array}{l} \theta^i=1\mbox{ for }\tau_+\\
\theta^i=\mbox{sg}(\frac{N+1}{2}-i)\mbox{ for }\tau_-
\mbox{ and }N=2n
\end{array}\right.
\label{tau}
\ee
The folded \cw-algebra is then defined by:
\begin{defi}
The folded $\cw_p(N)^\pm$ algebra is defined by the coset $\cw_p(N)/\cj$
where \cj\ is the ideal generated by $W^{ij}_n-\tau_\pm(W^{ij}_n)$.
\end{defi}
Note that \cj\ is an ideal for the product law, and one can show that the 
coset can be provided with the bracket of the $\cw_p(N)$ algebra (see
\cite{Ytw} for more detail). 

Now, one can prove \cite{fold,Ytw}:
\begin{theor}
$\cw_p(2n)^+$ (resp. $\cw_p(2n+1)^+$ and $p=2k+1$, resp. $\cw_p(2n)^-$) 
 is $\cw[so(2np),n.sl(p)]$ (resp. 
$\cw[so((2n+1)p),n.sl(p)\oplus so({k})]$, resp. $\cw[sp(2np),n.sl(p)]$). 
\end{theor}

\subsection{Twisted Yangians}
In the same way $\cu[gl(Np)]$ and $\cw_p(N)$ have been folded into
$\cu(\cg)$ and $\cw(\cg,\ch)$ with $\cg=so(M)$ and $sp(2M)$, one 
naturally considers the case of $Y(N)$. However, although Yangians based on
$so(M)$ and $sp(2M)$ exist, it is not these Hopf algebras that are obtained
through this procedure, but another type of algebras, named twisted 
Yangians \cite{defYtw}. More precisely, the automorphism (\ref{tau}) takes here 
the form 
\be
\tau(T(u))=T^t(-u)\ \mbox{ with }\ T^t(u)=\sum_{i,j} T^{ij}(u)E^t_{ij}
\ \mbox{ and }\ E^t_{ij}=\theta^{i}\theta^{j}E_{N+1-j,N+1-i}
\ee
It can be shown that $\tau$ is an automorphism of $Y(N)$. From this
automorphism, one defines
\be
S(u)=T(u)\tau(T(u))
\ee
Essentially, two classes of automorphisms appear, labeled 
by a parameter $\theta_{0}=\pm1$:
\be
\begin{array}{lll}
\mbox{ For }Y^+(N) \ : &\theta^{i}=1,\, \forall i & \ (\theta_0=1)\\
\mbox{ For }Y^-(2n) \ : &\theta^{i}=\mbox{sg}(\frac{N+1}{2}-i),\, \forall i & \ (\theta_0=-1)
\end{array} 
\ee

This defines a subalgebra $Y^\pm(N)$ of $Y(N)$, whose commutation relations are 
coded in
\be
R_{12}(u-v)\, S_{1}(u)\, R'_{12}(u+v)\, S_{2}(v) = 
S_{2}(v)\, R'_{12}(u+v)\, S_{1}(u)\, R_{12}(u-v)
\label{rsrs}
\ee
where $R(x)$ is given by (\ref{Rmat}), and
\bea
&&R'(x)=(\tau\otimes\II)(R(x))=(\II\otimes\tau)(R(x))=\II-\frac{1}{x}Q_{12}\nonu
&&\mbox{ with }Q_{12}=\sum_{i,j=1}^N \theta^{i}\theta^{j}
E_{ij}\otimes E_{N+1-i,N+1-j}
\eea
The finite dimensional irreducible representations and the center of 
$Y^\pm(N)$ have been determined in \cite{Mol}.

At the classical level,  $S(u)$
generates a Poisson subalgebra  $Y(N)$, the Poisson brackets 
being defined by
\[
\begin{array}{l}
\{S_1(u),S_2(v)\} = [r_{12}(u-v),S_1(u)S_2(v)]+
S_2(v)r'_{12}(u+v)S_1(u)  -S_1(u)r'_{12}(u+v)S_2(v)\\
\mbox{where } \ r'_{12}(x)=(\II\otimes\tau)r_{12}(x)=(\tau\otimes\II)
r_{12}(x)
\end{array}
\]
The level one generators of this subalgebra form the Lie algebra $so(M)$
or $sp(2M)$, but the total subalgebra is \underline{not} the Yangian
based on $so(M)$ or $sp(2M)$ (see \cite{MNO} for more details). 
However, it is these algebras
that are involved in the comparison with \cw-algebras:
\begin{theor}
The truncated classical Yangians $Y^\pm_p(N)$ are \cw-algebras. More precisely:
\be
\begin{array}{l}
Y_p(2n)^- \ \equiv\ \cw[sp(2np), n.sl(p)] \ \ ;\ \ 
Y_p(2n)^+ \ \equiv \ \cw[so(2np), n.sl(p)] \\
Y_p(2n+1)^+ \ \equiv\ \cw[so((2n+1)p), n.sl(p)\oplus so(k)]\ \mbox{ 
with }\ p=2k+1 
\end{array}
\ee
where $\equiv$ denotes algebra isomorphisms. The truncation is 
defined as in theorem \ref{theo-wrtt}. 
\end{theor}
Let us remark that, as in the case of $Y_p(N)$, the isomorphism cannot
be extended to a Hopf algebra isomorphism, the (untruncated)
twisted Yangians $Y(N)$ being even not Hopf algebras (only left 
coideals in $Y(N)$).

As for the Yangian $Y(N)$, this isomorphism provides a simple way of 
quantizing the \cw-algebras. One can also use it to determine
the center and the finite-dimensional irreducible representations 
of these \cw-algebras, see \cite{Ytw} for more details. 

\sect{Generalization to superalgebras\label{super}}
Once again, one can apply the same technics to the case of super-Yangians
and \cw-superalgebras. As for $Y(N)$ and $gl(N)$, the case of $gl(M|N)$ 
singles out.
\subsection{Super-Yangian $Y(M|N)$}
They are based on the superalgebra $gl(M|N)$ in the same way $Y(N)$ is based 
on $gl(N)$. They have been defined in \cite{defSY} and their 
representations are studied in \cite{zhang}.
One defines a $\ZZ_{2}$-grading
\be
[T^{ij}_{(n)}]=[i]+[j]\ \mbox{ with }\ \left\{
\begin{array}{l}
    {[i]}=0\mbox{ for } 1\leq i\leq M\\ 
    {[i]}=1\mbox{ for } M+1\leq i\leq M+N
 \end{array}
\right.
\ee
and introduces as usual:
\be
T(u)=\sum_{i,j=1}^{M+N}\sum_{n\geq 0}u^{-n}\, T^{ij}_{(n)}E_{ij}
=\sum_{i,j=1}^{M+N}\, T^{ij}(u)E_{ij}
\ \mbox{ and }\ P_{12}=\sum_{i,j}(-1)^{[i][j]} E_{ij}\otimes E_{ji}
\ee
The super-Yangian is then defined by
\be
 R_{12}(u-v)T_1(u)T_2(v)=T_2(v)T_1(u)R_{12}(u-v)
 \ \mbox{ with }\ R_{12}(u)=\II-\frac{1}{u}P_{12}
\ee
where we have introduced graded tensor products:
\be
T_{1}(u)= \sum_{i,j,k,l}(-1)^{([i]+[j])[k]}\, T_{ij}(u)\delta_{kl}\ E_{ij}\otimes E_{kl}
\ \mbox{ and }\
T_{2}(u)= \sum_{i,j} T_{ij}(u) \ \II\otimes E_{ij}
\ee
It is a graded Hopf algebra, and its $R$-matrix obeys a graded Yang-Baxter algebra. 
Their classical version is defined as in section \ref{class}.
We refer to \cite{defSY,zhang,sYW} for more details.
\subsection{$\cw(M|N)$-superalgebras}
Starting from the superalgebra $gl(M|N)$ and using $sl(2)$ embeddings, one can construct 
\cw-superalgebras. 
The $sl(2)$ generators being bosonic, they belong to the $gl(M)\oplus 
gl(N)$ subalgebra, and the procedure is the same as in section 
\ref{Walg}. The only difference comes with the fermionic generators which 
have to be constrained to Grassmann constant for consistency: see 
\cite{classW} for details. 

As for the $gl(N)$ case, one selects a special class of \cw-superalgebras: 
the finite \cw-superalgebras $\cw_{p}(M|N)=\cw[gl(pM|pN),(M+N).sl(p)]$, 
where $(M+N).sl(p)$ denotes the direct sum of 
$(M+N)$ algebras $sl(p)$, $M$ of them being included in the $gl(M)$ 
subalgebra of $gl(M|N)$ and the $N$ remaining in its $gl(N)$ subalgebra.
Then, one can prove
\begin{theor}
    The $\cw_{p}(M|N)$ superalgebras are isomorphic to the truncation 
    at level $p$ of the classical super-Yangian $Y(M|N)$.
\end{theor}
This isomorphism allows us to classify the irreducible 
finite-dimensional representations of the $\cw_{p}(M|N)$ 
superalgebras \cite{sYW}.

\subsection{Twisted super-Yangians}
Similarly to the twisted Yangians, one can define the twisting of 
super-Yangians\cite{sYtw}. This leads to subalgebras of $Y(M|N)$ which contains 
the orthosymplectic superalgebras. Mimicking the case of twisted Yangian, one introduces 
an automorphism of $Y(M|N)$:
\be
\begin{array}{l}
\tau(T^{ij}(u))=(-1)^{[i]([j]+1)}\theta_{i}\theta_{j}T^{\barj\bari}(-u)
\ \mbox{ with }\ \theta_{i}=\pm1\ \mbox{ ; }\ 
(-1)^{[i]}\theta_{i}\theta_{\bari}=\theta_{0}=\pm1 \\
\bari=M+1-i\mbox{ for }1\leq i\leq M\\
\bari=2M+N+1-i\mbox{ for }M+1\leq i\leq M+N
\end{array}
\label{supertau}
\ee
However, the presence of fermionic generators forces to 
have\footnote{Up to the Hopf algebra isomorphism $Y(M|N)\leftrightarrow 
Y(N|M)$ which identifies $M=2m$ and $\theta_0=-1$ with $N=2n$ and $\theta_0=1$, 
see \cite{sYtw}.} $N=2n$ 
and $\theta_{0}=+1$. Thus, one is led to the definition:
\begin{defi}
    The twisted Yangian $Y(M|2n)^{+}$ is the subalgebra of $Y(M|2n)$ 
    generated by $S(u)=T(u)\tau(T(u))$, where $\tau$ is defined in (\ref{supertau}) 
    with
    \be
\begin{array}{l}
\theta_{i}=1\ \mbox{ for }1\leq i\leq M\\
\theta_{i}=\mbox{sg}(\frac{2M+N+1}{2}-i)\ \mbox{ for }M+1\leq i\leq M+2n
\end{array} 
\ee
\end{defi}
{From} this definition, one proves that $S(u)$ obeys the rules:
\bea
&&R_{12}(u-v)\, S_{1}(u)\, R'_{12}(u+v)\, S_{2}(v) = 
S_{2}(v)\, R'_{12}(u+v)\, S_{1}(u)\, R_{12}(u-v)\\
&&\tau(S(u))=S(-u)+\frac{1}{2u}\left(S(u)-S(-u)\right)
\eea
The irreducible finite-dimensional representations of the twisted 
super-Yangian are studied in \cite{sYtw}.

\null

As far as \cw-superalgebras are concerned, their folding have been 
introduced in \cite{fold} and shown to lead to \cw-superalgebras based 
on $osp(M|N)$ superalgebras. Considering a special class of folded 
\cw-superalgebras, one gets again:
\begin{theor}
    Let $\cw_{p}(M|2n)^+$ be the  
    $\cw[osp(Mp|2np),(\left[\frac{M}{2}\right]+n)sl(p)\oplus\eps_M\, 
    so(k)]$ 
    superalgebra, where $\eps_M\equiv M$ $[${\rm mod} $2]$ and $p$ is  
    chosen odd ($p=2k+1$) when $M$ is odd. $\cw_{p}(M|2n)^+$ is
    isomorphic to the truncation 
    at level $p$ of the twisted super-Yangian $Y(M|2n)^+$.
\end{theor}
It allows to classify the finite dimensional 
representations of the $\cw_{p}(M|2n)^+$ algebras \cite{sYtw}.

\sect{Conclusion\label{concl}}
A wide class of \cw-(super)algebras are shown to be isomorphic to the 
truncation of (super)(twisted) Yangians. This isomorphism allows to 
classify all the irreducible finite-dimensional representations of 
these \cw-algebras. 

Moreover, since there are many more \cw-algebras, the
connection let us hope that a generalization of Yangians (as Hopf algebras)
is available. The same is valid for affine \cw-algebras, which should lead to
two-parameters generalization of Yangians.

Finally, the application to physical models, such as Nonlinear Schr\"odinger
equation has to be studied.

\end{document}